\documentclass[preprint,endfloats*]{revtex4}
\nonstopmode

\usepackage{amsmath}
\usepackage{amsbsy}

\include{dirnot}

\newcommand{\bs}{\boldsymbol}

\begin{document}

\title{Reply to ``Comment on 'Critique of the foundations of time-dependent density 
functional theory' ''}
\author{J. Schirmer}
\affiliation{Theoretische Chemie, Physikalisch-Chemisches Institut,
Universit\"at Heidelberg, \\
D-69120 Heidelberg, Germany}
\author{A. Dreuw}
\affiliation{Institut f\"ur Physikalische und Theoretische Chemie,
Universit\"at Frankfurt, Germany\\
D-60439 Frankfurt}
\date{\today}

\begin{abstract}
The Comment by Holas \textit{et al.} [A. Holas, M. Cinal, and N. H. March,
Phys. Rev. A \textbf{78}, 016501 (2008)] on our recent paper 
[J. Schirmer and A. Dreuw, Phys. Rev. A \textbf{75}, 022513 (2007)].
is an appropriate and valuable contribution. As a small addendum we   
briefly comment on the relationship between the radical Kohn-Sham (rKS)
form of density-functional theory and previous one-electron (particle) potential 
(OPP) developments.
\end{abstract}
\maketitle
\newpage

In a Comment, Holas \emph{et al.}~\cite{hol08:016501} address two distinct topics
related to Secs.~3 and 4, respectively, of our recent
paper~\cite{sch07:022513}, henceforth referred to as I. Essentially, we appreciate
their contribution as  
an appropriate and valuable addendum.  
Moreover, the Comment is a welcome opportunity to
rectify a regrettable omission, namely the failure to cite
two previous papers, Refs.~(2) and (9) of the Comment, 
which are clearly of relevance in the context of our paper. 

However, we would like to comment briefly on the reference to the so-called
one-electron potentials (cf. beginning of the third paragraph, Refs.~(3-6), and 
the final paragraph of the Comment). In spite of some apparent similarity due to
the use of the square root of the density function, $\sqrt{\rho(\bs{r})}$, there is a
major difference between the one-electron (particle) potential (OPP) 
approach~\cite{hun75:237,hun86:197}
(see also
Lassettre~\cite{las85:1709}, Kohout~\cite{koh02:12}, and references therein)
and the radical Kohn-Sham (rKS) form of density-functional theory (DFT).     
The OPP approach is a (rather trivial) factorization of the full $N$-electron
wave function, say for the ground state, according to 
\begin{equation}
\label{eq:oep}
\Psi_0(\bs{r} s,\bs{r}_2 s_2, \dots,\bs{r}_N s_N) = \phi (\bs{r})\, 
              \Theta (\bs{r} s,\bs{r}_2 s_2, \dots,\bs{r}_N s_N) 
\end{equation}
where $\phi (\bs{r})= N^{-\frac{1}{2}} \sqrt{\rho_0(\bs{r})}$ and 
$\Theta(\bs{r} s,\bs{r}_2 s_2, \dots,\bs{r}_N s_N)  
= \phi (\bs{r})^{-1} \Psi_0(\bs{r} s,\bs{r}_2 s_2, \dots,\bs{r}_N s_N)$.
The full Schr\"odinger equation (SE) can then be transformed into
a one-electron Schr\"odinger type equation for the ``orbital'' $\phi (\bs{r})$
with a local one-electron potential, $v_{OPP}(\bs{r})$, obtained by inserting the 
ansatz~(\ref{eq:oep}) into the full SE and integrating over all spatial and
spin degrees of freedom except for $\bs{r}$ (see Refs.~\cite{hun86:197}, 
\cite{las85:1709}, and~\cite{bui89:4190}). 
It should be clear that this procedure is not designed as a method to determine
the (exact) density function. The idea is that the $v_{OPP}(\bs{r})$ potential,
which can only be constructed when the full $N$-electron wave function is 
available, has some benefit  
to characterizing the system under consideration at the level of a 3-d spatial potential.

While rKS theory also operates at the one-particle level,
the key ingredient here (as in the usual KS variant) is not a potential but 
a potential-functional (PF), $\tilde{v}^{eff}[\rho](\bs{r})$,
being of the form (see Eq.~39 in I) 
\begin{equation} 
\tilde{v}^{eff}[\rho](\bs{r}) = v(\bs{r}) + J[\rho](\bs{r}) + \tilde{v}_{xc}[\rho](\bs{r})
\end{equation}
The use of this exact (or approximate) PF in the one-orbital rKS equation establishes a   
fixed-point iteration scheme (FPI), allowing one to determine the 
exact (or approximate) ground-state density function $\rho_0(\bs{r})$ as the fixed-point.
The potential associated with $\rho_0(\bs{r})$, referred to as the (exact) rKS potential,
\begin{equation}
v_{KS}(\bs{r}) =  \tilde{v}^{eff}[\rho_0](\bs{r}) 
\end{equation}
can readily be be obtained from $\rho_0(\bs{r})$ according to (see Eq.~39 in I)
\begin{equation}
\label{eq:inv}
v_{KS}(\bs{r}) = \frac{\nabla^2 \sqrt{\rho_0(\bs{r})}}{2 \sqrt{\rho_0(\bs{r})}} + \epsilon
\end{equation}  
by an obvious ``inversion'' of the rKS equation at the fixed-point. Here the 
rKS and OPP schemes make contact, since both  
give rise to the same orbital, $\phi(\bs{r}) = \sqrt{\rho_0(\bs{r})}$.
This means that $v_{OPP}(\bs{r})$ and $v_{KS}(\bs{r})$ 
are identical up to a constant. Obviously, the inversion procedure~(\ref{eq:inv}) offers
a much simpler way to construct the OPP than the original approach.  
One may ponder whether the (exact) rKS potential or the OPP is of particular 
significance. While it certainly does not contain more physical information 
than the corresponding density function, the manifestation of this 
information in the shape of a local one-particle potential has, perhaps, some
descriptive value.

As we have mentioned in I, the rKS formulation of DFT is not new. This fully 
justified simplification of the usual ($N$-electron) KS theory has been proposed
previously by Levy \emph{et al}.~\cite{lev84:2745} (Ref.~54 in I) and
by Holas and March~\cite{hol91:5521} (Ref.~9 of Comment), independently. 
In view of the clarification 
given above, we do not feel that the OPP developments should rank
among the legitimate predecessors of the rKS approach.

\end{document}